# DMap: A Distributed Blockchain-based Framework for Online Mapping in Smart City

Fatemeh MohammadZadeh

Department of Computer Engineering

University of Guilan

Guilan,Iran

mohammadzadeh.fatemeh46@gmail.com

Seyed Ali Mirghasemi

Department of Computer Engineering

University of Guilan

Guilan,Iran

Ali.mirghasemi.94@gmail.com

Ali Dorri

School of Computer Science and Engineering

The University of New South Wales

Sydney, Australia

ali.dorri@unsw.edu.au

HamidReza Ahmadifar

Department of Computer Engineering

University of Guilan

Guilan,Iran

ahmadifar87@gmail.com

*Abstract* - **Smart cities are growing significantly due to the growth of smart connected vehicles and Internet of Things (IoT) where a wide range of devices are connected to share data. Online mapping is one of the fundamental services offered in smart cities which enables the vehicle owners to find shortest or fastest direction toward a destination thus reduces travel cost and air pollution. However, existing online mapping services rely on centralized servers, e.g., Google, which collect data from users to offer service. This method is unlikely to scale with growth in the smart city participants and introduces privacy concerns and data wall where data of the users is managed by big companies. To address these challenges, in this paper we introduce DMap, a blockchain-based platform where the users can share data in an anonymous distributed manner with service providers. To the best of our knowledge, DMap is the first distributed blockchain-based solution for online mapping. To improve the scalability of the blockchain, we propose to use edge-computing along with blockchain. To protect against malicious vehicles that may inject false data, we define a reputation system where the collected data of the vehicles is verified by monitoring the neighbor data. We introduce data marketplace where users can trade their data to address data wall challenge.**

## INTRODUCTION

In recent years, the vehicles are getting connected to other vehicles and smart infrastructures in smart cities and generally to the Internet, thus become part of the Internet of Things (IoT). Smart vehicles are equipped with a wide range of devices and built-in sensors like cameras, radar sensors, and built-in GPS devices. The collected data by the smart vehicles is expected to reach 4TB per day [1] by 2020. On one hand, the data collected can be used by service providers to offer a wide range of personalized sophisticated services, e.g., online mapping service, congestion control, and locating parking spots. On the other hand, the service providers can build virtual profiles of the users' activities and thus compromise their privacy. The SPs may also trade the data of the users without their consent which also comprises the user privacy.

The accuracy of the received services is critical for the smart vehicle owners. The services offered by the SPs rely on the collected data from the users, thus the SPs with more customers will provide more accurate services to their users which leads to attracting even more users. This lead to "data wall" challenge where the data of the smart vehicles is owned and managed by large companies. Smaller companies cannot compete with large companies due to lack of real-time data from users.

Location-Based Services (LBS) are among the main services offered using the data collected by the smart vehicles due to the vehicle mobility. Using LBS the SPs monitor real-time location of the vehicles and offer personalizes services, including locating nearest parking spot and online mapping which eventually reduces the travel time and air pollution and cost. The city managers also benefit from LBS to control the traffic and congestion and plan for city management projects such as train route, new road, and traffic lights. Although SPs can utilize the location data to offer personalized services to the user, they can virtually build a profile of user activities and track the user location which in turn compromises the user privacy.

As noted earlier, the services offer by SPs largely relies on accurate data which made data unprecedentedly important in smart cities. Many data-driven companies trade the data of their users which leads to the introduction of an ever-growing technology known as data market place. The data produced by the end-users or their associated devices is fed to the data market place, where the user can manage all his data and trade with SPs or other users in return for receiving incentives that is mainly monetary. It is critical that the users have control over their data and be able to audit any exchange of their data. The market place users need to trust to the data generator and devices do not generate false data, i.e., they need to trust that the received data is correct. However, existing market places generally assume that the devices are secure and thus generate trusted data, while this might not always be correct. Thus, data market place demands a method to achieve trust to the data itself.

In recent years, blockchain has attracted a significant attention from both academia and practitioners as a means to provide decentralization, anonymity, and security. Blockchain is an immutable timestamp ledger of blocks that is used for storing data. The participating nodes in blockchain use transactions to communicate which are cryptographically signed. All transactions are broadcast to the network and verified by all participating nodes without relying on a central authority or broker. Particular nodes in the network, known as miners or verifiers, collect transactions in form of blocks and append in the blockchain. The latter involves following a consensus algorithm which ensures blockchain security against malicious miners that try to store invalid transactions in the blockchain. As shown in Figure 1, each block maintains the hash of the previous block which ensures blockchain immutability. Changing the content of a block changes the corresponding hash in the subsequent blocks which will reveal the attack. Blockchain users use changeable keys to protect their identity which introduces high level of anonymity.

The main contribution of this paper is to present DMap, a distributed blockchain-based solution for online mapping which to the best of our knowledge is the first work in this area. DMap uses a public blockchain where any node can join the blockchain and generate transactions. The participating nodes, that can be smart vehicles, Road Side Infrastructures (RSI), and city managers, are known by a changeable PK which introduces anonymity. To increase scalability, only the high-resource available participants, e.g., RSIs, and city managers manage the blockchain by generating blocks and verifying transactions. The smart vehicle owners can trade the data of their vehicles using blockchain with other participants in a distributed and anonymous manner. Recall that trust to the generated data by the devices is challenging in existing market places. To address this challenge, we propose to employ edge-computing, where RSI collects data of smart vehicles and validates the data as the edge of the network. To validate data the RSI examines the received data from each vehicle with the data received by other vehicles in the same region. If the RSI detects inconsistency between the data produced by one vehicle compared to other surrounding vehicles, it marks the data as untrusted and discards the transaction. The RSI then signs the transactions and broadcast to the blockchain. It is assumed that each RSI uses a predefined key which is certified by a Certificate Authority (CA) and these keys are known to all blockchain miners used for verifying the data. By employing edge-computing DMap increases scalability and reduces processing overhead in blockchain to ensure trust in data sources. We also used cloud service as automotive data market place to store verified data.

DMap enables blockchain participants to request the traffic data in a particular region that can be used either for city planning or online mapping. For this, the participants send their request to a "rule table" which is essentially an API in the cloud storage that connects the cloud to the blockchain similar to wallet softwares in conventional blockchains. The data of the smart vehicles is classified based on the location of the corresponding RSI. To conduct the trading, DMap employs smart contract which determines how long the user is allowed to use the data and to which data the user is permitted to access.

The rest of the paper is structured as follows: in Section II we describe related work. Details of DMap are discussed in Section III. Section IV describes data trade process using DMap. The security of DMap is analyzed in Section V and finally, in section VI we conclude the paper.

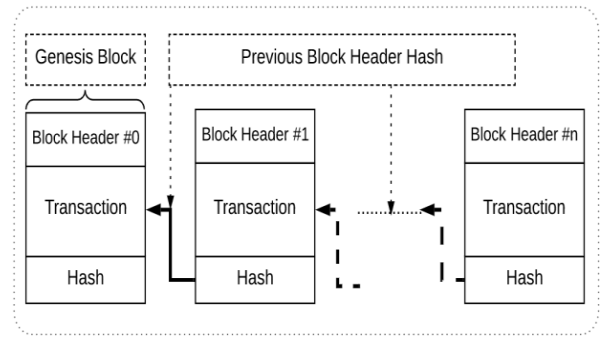

Figure 1. The structure of the Blockchain

## II. RELATED WORK

Blockchain applications in smart connected vehicles and smart cities have received a tremendous attention in recent years. In this section, we propose a literature review of the relevant works to DMap.

Authors in [7] proposed a blockchain-based framework known as speedychain which is adopted for smart city scenarios. This framework uses permissioned blockchain to allow smart vehicles transmit data in a private and scalable manner while using a two-level trust model in which vehicles can trust RSIs and vice versa. The proposed method reduces latency in transferring data by storing transactions corresponding to each RSI in separate blocks.

Authors in [6] presented a blockchain-based solution for automotive platform to overcome many traditional vehicular challenges including privacy preservation of the user while guaranteeing the security of the vehicle. In this approach each vehicle is equipped with an in-vehicle storage which is responsible for gathering privacy sensitive data which also prevents privacy leakage. This architecture also provides automotive services update without the interference of a central authority.

Authors in [8] set up a blockchain-based solution to reduce the self-driving vehicle accident casualties with the capability to verify and store the accident scenarios where cell-based vehicular network oversees accident data of other vehicles. They presented a new proof of event mechanism to ensure the accuracy of the car accident data and its trustworthiness which are recorded in to the block of transactions. In the proposed architecture vehicles that were directly involved in the accident scene generate an accident transaction through vehicular network. They also presented a dynamic federation consensus model to verify the block of events.

Authors in [9] have adopted the combination of vehicular communication technology and cloud computing together with blockchain technology to deploy an intelligent data sharing platform. Their architecture aimed to provide the privacy of Intelligent Vehicle. This architecture presents seven-layer vehicle communication framework while meeting the increasingly strict security requirements needs.

## III- D-Map: A Distributed solution to online mapping

This section describes the details of the proposed architecture. Participating nodes in the network are smart vehicle, city manager, service providers (SPs), and Road Side Infrastructures (RSIs) which join a public blockchain. To improve scalability and reduce the overhead in smart vehicles, the blockchain is managed by high resource available devices including RSI, SPs, and city managers. The smart vehicles connect to the RSI in their region to receive service from the blockchain. Note that the smart vehicles are mobile entities and thus will need to change their associated RSI as they move. We assume soft hand over methods as in [6] are in place to manage the vehicle movement between RSIs.

The smart vehicles share their location data with RSIs. Communication between RSI and the vehicles occur through transaction. RSIs are registered entities in a Certificate Authority (CA) and thus their key is known to all blockchain network. We assume that RSIs are trusted entities. A vehicle is a mobile node which is identified by a PK[1]. Every vehicle will interact with a RSI which is located in its close proximity. Every RSI collects information from vehicles in its neighborhood and will prove the transaction during verification process. Thereafter, RSI will send the transaction to the blockchain with its own certificate which essentially is the RSI signature, so that miners can verify that transaction has not tempered with. The valid transactions are then mined in the blockchain.

Smart vehicles produce a huge volume of data which contains privacy-sensitive information about the vehicle owner. Thus, it is critical for the vehicle owners to be able to control which data of the vehicle to share with requesters. Mapping services utilize the location data of the vehicle, and thus should not be able to use other data of the vehicle. We assume that mechanism such as the one used in [4] would be effective, thereby the driver can manage data sharing and decides on what type of data to share.

All transactions contain the signature of data generator, i.e., the smart vehicle, and data validator which is RSI. Based on the number of signatures in transactions, they are divided into two categories:

- Single signature: this type of transaction requires only one signature to be considered as a valid transaction.
- Multisign[2]: this type of transaction requires *m* out of *n* nodes to sign the transactions to be considered as a valid transaction in blockchain.

Each vehicle populates a single signature transaction (1) that contains its geographical location data and communicates with nearest RSI nodes through any short-range communication technology e.g. Wi-Fi. The vehicle uses a fresh PK to generate new transactions and also ensures annoniminity. This prevents malicious nodes to track the vehicle location, which in turn compromises the privacy of the vehicle owner. The structure of the transaction sent by the vehicle to the RSI is as below:

$$T_{data} = [loc||event||timestamp||P_K||Vehicle_{sign}] \quad (1)$$

Where "*loc*" is the location where an event has occurred or the data has been collected, "*event*" is the type of situation that occurred on the road e.g. road damage or parking spots. *"timestamp"* is the information which identifies when a certain event has occurred. Finally, the vehicle populates its PK and the corresponding signature in the last two fields.

The vehicles send the transactions to their associated RSI. The RSI needs to verify that the generated data is accurate and trusted. The RSIs employ neighbor monitoring methods such as in [10] for this aim. When RSI receives data from a vehicle, it must wait to receive data from other vehicles. If the data received from a vehicle is aligned with the data received from other vehicles, the RSI can trust to the data and sign the transactions as valid. Otherwise the transaction will be considered as invalid and is discarded. Note that as the vehicles are in close proximity, it is expected that the data from the vehicles have overlap and thus can be used to verify the data of other vehicles. In case the vehicles are not in close proximity, the RSI will set a flag in transaction indicating that the data might not be trusted. If the RSI set the flag value as 1 meaning that the transaction is trustworthy, otherwise it should be rejected by the blockchain members.

---

[1] Public key

[2] Multi-signature

Once the RSI verified all received transactions, it generates a transaction which is structured as below:

$$T_{RSI} = [RSI_{sign}||T_{data}||vehicles_{sign}||vehicles_{pk}||flag] \quad (2)$$

The "$RSI_{sign}$" is an identifier for the RSI while $T_{data}$ is the data of sent by the vehicles. And the next two fields are vehicles signature and the corresponding PK respectively. Using this transaction, the participating nodes in the blockchain, can ensure that the data is trusted and verified by the RSI. It is noteworthy that since the data obtained from several vehicles may be similar and accurate, RSI will send only one copy of the data along with the signatures of the data producers in the multisign transaction shown above. The last field is the flag which is set by the RSI which identifies the data trustworthiness (as explained earlier).

In case that the verification is completely proved by the RSI the multisign transaction is broadcast to blockchain. It is noteworthy that in DMap all multisign transactions generated by each RSI are organized in a separate ledger which potentially protects against Sybil attack where a node floods the network with fake transactions.

## IV. BLOCKCHAIN SMART CONTRACT

In DMap, the exchange of the data can be triggered by either the smart vehicle owner or a SP. The smart vehicle owners can share their data with other users, including online map SPs, e.g., Google, using smart contracts. They can share the real-time or previously stored data for a particular time-frame requested by the SP. To do so, the vehicle owner stores a smart contract in blockchain that contains the PK of the SP, and the timespan in which the SP is allowed to access the data. SPs can request the data of all vehicles in a particular area. This in particular enables the SP to collect necessary information for online mapping service. To do so, the SP sends a data request transaction to all participants connected to a particular RSI or group of RSIs. The data request transaction contains the geographic coordinates of the requested area and the period of time in which the SP needs to access the data.

When a RSI is certified by the CA, a dedicated storage space is created for the RSI in the cloud storage that is referred to as "data directory". The cloud storage is employed to store the data while data access control is managed through the blockchain. The data directory facilitates retrieving data as the data is technically grouped based on the location.

As shown in figure 2, the cloud storage is connected to the blockchain through a "rule table" API which facilitates the store and retrieve actions by the blockchain participants. As discussed earlier the rule table is placed in to the cloud storage. Recall from Section III that vehicles broadcast their single-signature transactions to Blockchain (step1 in figure 4). To store data in the cloud, the RSI unicasts data to the cloud along with the corresponding transaction (step 2 in figure 4). The rule table fetches the RSI key in the transaction and stores the data in the dedicated storage space to that particular RSI (step4 in figure 4). To retrieve data, the rule table first ensures that the blockchain participant that requests data has permission to access the data (as discussed later in this section) (step6 in figure 4). The data requester generates an access transaction, which is a multisign transaction and contains either the address of a smart contract where the data owner grants permission to access the data, or the signature of the data owner (step 5 in figure 4). The requester and the data owner must agree on the price of the data beforehand. If requester has permission to access the data, the rule table signs the corresponding access transaction and directly sends the data to the requester (step 7 and 8 in figure 4). Note that the rule table retrieves data based on the location of the data and its corresponding RSI. The signed access transaction is then stored in the blockchain (step 9 in figure 4). The rule table enables the blockchain participants to explore availability of particular data and volume of available data. Detail of the proposed architecture is outlined in figure 3. To reduce the associated overheads for retrieving data, DMap employs Elastic search [11] which enables marketplace to be able to support any standardized JSON format for location data.

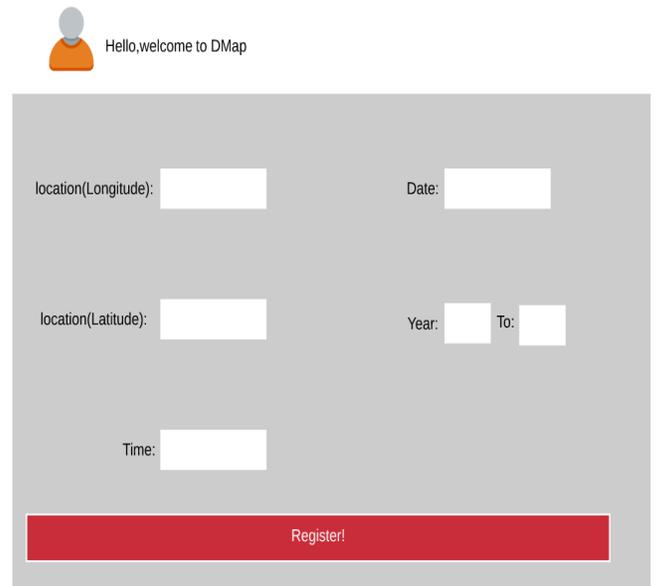

Figure 2. An identical view to find location data

In this section we analyze the privacy and security of the proposed architecture.

*Privacy:* In DMap, each smart vehicle employs a changeable PK as its identity which in turn introduces high-level of anonymity. There is no link between the PK of the vehicle and the real-identity of the user. In DMap the users have full control on their data and are able to audit the accesses to their data.

The malicious nodes may attempt to deanonymize a user by linking multiple identities to a single user which is known as linking attack in literature. In DMap, each transaction is generated using a fresh PK which prevents this attack. The trust to the data generated by the vehicles is later achieved using neighbor monitoring.

In order to protect the RSIs safe against malicious vehicles that may produces false data, they start neighbor monitoring meaning that, each time it compares the transactions transmitted by each vehicle with the transactions of other vehicles a vehicle whose data varies by data from other vehicles is known as a malicious node and its data is rejected by the RSI.

*Security:* The security of DMap is inherited from the security of blockchain where all transactions are encrypted and signed by the transaction generator. The data of the vehicle is stored in a cloud storage, where the data owner manages the access control using blockchain. All interactions by participants are recorded in the blockchain which introduces high auditability and transparency. Thus, the vehicle owner can detect any unauthorized access to his data.

A malicious vehicle may try to falsify the network or impact online mapping services by injecting fake data. To prevent this attack, in DMap the RSIs conduce neighbor monitoring. In this method, the RSI compares the data collected by all RSIs and trust the most common data as the valid data. Thus, the fake data injected by a malicious node can be detected. However, if the number of malicious nodes in an area exceeds the number of honest nodes, the data might be affected, and the malicious nodes may be able to inject fake data. Note that given the large number of smart connected vehicles in an area, it is hard for the attacker to compromise the majority of the vehicles.

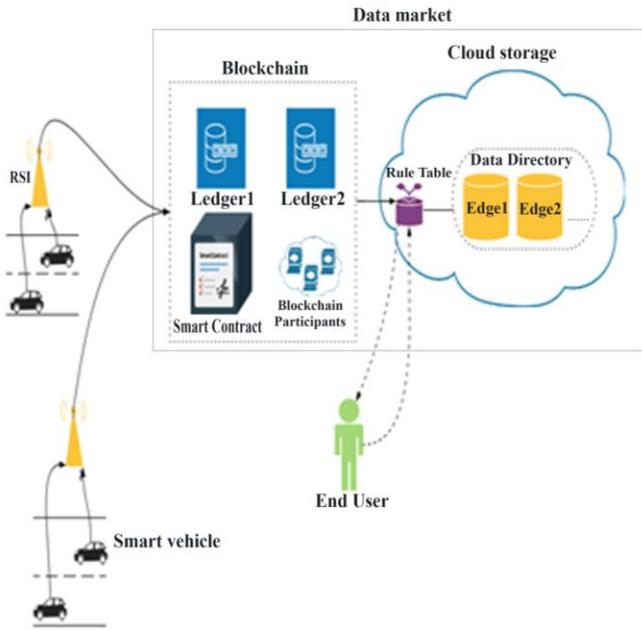

Figure 3. Details of the proposed architecture

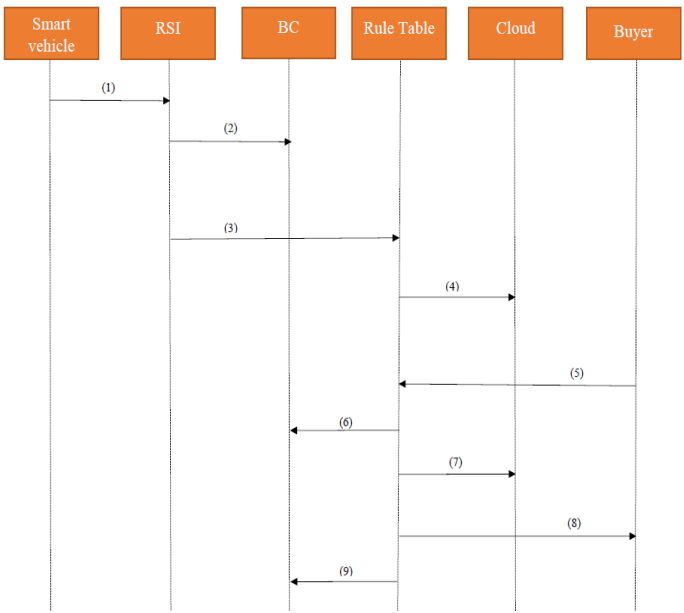

Figure 4. The process of intractions between entities

## V. Privacy And Security Analysis

## VI. Conclusion

Smart vehicles generate significant volume of valuable data that can be used by the smart vehicle service providers to offer personalized services. In this paper, we proposed DMap, a decentralized blockchain-based framework that enables the vehicle owners to monetarize the data generated by their vehicle. DMap enables location-based service providers to purchase data from the users in real-time. DMap uses edge computing to enhance the scalability and reduce the processing overhead on participating nodes. To protect against malicious vehicles that inject fake data, the RSI validates data received by the vehicles by employing neighbor monitoring methods.